\documentclass[12pt]{article}
\usepackage{graphicx}
\usepackage{amsmath,amssymb,amsfonts}

\begin{document}

\title{Electromagnetic waves in the gravitational field of massive dark halos}

\author{Shahen Hacyan
}

\renewcommand{\theequation}{\arabic{section}.\arabic{equation}}

\maketitle
\begin{center}

{\it  Instituto de F\'{\i}sica,} {\it Universidad Nacional Aut\'onoma de M\'exico,}

{\it A. P. 20-364, M\'exico D. F. 01000, Mexico.}

\end{center}
\vskip0.5cm

\begin{abstract}

The propagation of plane electromagnetic waves in the gravitational field inside a rotating cloud of dark matter
is analyzed. Formulas for the deflection and absorption of light,  and the rotation of the polarization plane are
obtained in closed form in terms of the mass density and the angular velocity of the cloud. It is shown that the
formulas can be considerably simplified for axisymmetric configurations. As an example, the formalism is applied
to a rotating massive cloud described by a Plummer potential.

\end{abstract}
e-mail: hacyan@fisica.unam.mx



 \maketitle

\newpage

\noindent{\it In memory of Jerzy Pleba\'{n}ski}

\section{Introduction}

The propagation of electromagnetic waves in a gravitational field is an interesting problem which has attracted
the attention of many authors since the late fifties \cite{skr,bal,pleb} and until more recently (see, e.g.,
\cite{ks,ser,nag}). Various approaches have been used to describe the physical interaction of light with gravity.
One of these approaches, based on the formal equivalence between a gravitational field and an anisotropic medium,
was used by Plebanski \cite{pleb} in a pioneering article of 1959 describing the scattering of electromagnetic
waves by massive bodies in motion. Using the weak field and short wave-length approximations, Plebanski obtained
formulas for the deflection angle and the polarization of the wave. As an illustration of the formalism, he worked
out the case of a rigidly rotating homogeneous sphere. Though such a model is rather unrealistic, the results
strongly suggested that the polarization plane of an electromagnetic wave is not rotated by the gravitational
field of a massive body unless the wave propagates through it. At that time, this result seemed to be of academic
interest only, since a material medium should affect the polarization of light in many ways that completely
overwhelm the effects of gravity. However, it is now usually accepted that large amounts of dark matter are
present in galactic haloes, possibly made of particles that do not interact electromagnetically with matter and
which manifest themselves through gravity only. Accordingly, the propagation of light in these massive
configurations is no longer a problem of academic interest and merits further investigations.

The first aim of the present paper, as presented in Section 2, is to simplify the formalism developed by Plebanski
\cite{pleb} and bring it to a form which is suitable for numerical calculations. The only assumption made is that
the gravitational field is stationary (for the interaction with gravitational waves, see, e.g., Ref. \cite{hac}).
The basic formulas for light deflection and rotation of polarization plane are obtained in closed forms directly
in terms of the mass density and velocity distribution of matter, instead of the gravitational potentials which
are usually difficult to calculate. A formula for the deflection of light in the geometrical optics approximation
is obtained in Section 3. Vectorial optics is considered in Section 4, where it is shown how the gravitational
field of a rotating mass distribution affects the polarization of light passing through it. In Section 5, it is
shown that the basic formulas can be simplified if the mass configuration has axial symmetry. As a particular
example, the results are applied to a model of a rigidly rotating cloud with a mass distribution given by a
Plummer potential \cite{plum}.

\section{Electromagnetic field}

Consider a space-time described by a metric of the form
\begin{equation}
g_{\alpha \beta} = \eta_{\alpha \beta} + \Delta g_{\alpha \beta},\label{delt-g}
\end{equation}
where $\Delta g_{\alpha \beta}$ is a small correction to the Minkowski metric $\eta_{\alpha \beta}$ in Cartesian
coordinates. According to the Einstein-Infeld-Hoffmann formalism \cite{EIH}, a distribution of matter with mass
density $\rho({\bf r},t)$ and local velocity ${\bf v} ({\bf r},t)$ produces, to lowest order, a deformation of the
space-time metric of the form
$$
\Delta g_{00} = - \frac{2 G}{c^2} \int dV' \frac{\rho({\bf r'},t)}{|{\bf r}-{\bf r'}|}
$$
$$
\Delta g_{0a} =  \frac{4 G}{c^3} \int dV' \frac{\rho({\bf r'},t) v_a ({\bf r'},t)}{|{\bf r}-{\bf r'}|}
$$
\begin{equation}
\Delta g_{ab} = - \delta_{ab} (1 - \Delta g_{00}) \label{gab}
\end{equation}
(we use signature $\{+---\}$).

Given the metric \eqref{delt-g}, the Maxwell equations in vacuum
\begin{equation}
F^{\alpha \beta}_{~~;\beta} =0, \quad  F_{[\alpha \beta , \gamma]}=0
\end{equation}
can be written as equations in a medium \cite{pleb}:
\begin{equation}
\frac{\partial}{\partial t} \big[(1 - \Delta g) {\bf E} + {\bf \Delta g} \times {\bf H} \big] - c\nabla \times
{\bf H} = {\bf 0}
\end{equation}
\begin{equation}
\nabla \cdot \big[ (1- \Delta g) {\bf E} + {\bf \Delta g} \times {\bf H}\big] = 0
\end{equation}
\begin{equation}
\frac{\partial}{\partial t} \big[ (1 - \Delta g) {\bf H} - {\bf \Delta g} \times {\bf E}\big] + c\nabla \times
{\bf E} = {\bf 0}
\end{equation}
\begin{equation}
\nabla \cdot \big[ (1- \Delta g) {\bf H} - {\bf \Delta g} \times {\bf E}\big] = 0 ,
\end{equation}
where $E_a = F_{a0}$ and $H_a = \frac{1}{2} (-g)^{1/2} \epsilon_{abc} F^{bc}$;  $\Delta g \equiv \Delta g_{00}$
and ${\bf \Delta g} \equiv \Delta g_{0a}$; $g$ is the determinant of $g_{\alpha \beta}$ and $\epsilon_{abc}$ is
the three-dimensional Levi-Civita symbol.

In the short wave-length approximation, the electromagnetic field vectors are assumed to be of the form
\begin{equation}
{\bf E} = ( {\bf E}^{(0)} + {\bf \Delta E} ) e^{i S},
\end{equation}
\begin{equation}
{\bf H} = ( {\bf H}^{(0)} + {\bf \Delta H} ) e^{i S},
\end{equation}
where $S$ is the eikonal function satisfying the equation
\begin{equation}
g^{\alpha \beta} S_{,\alpha} S_{,\beta} =0 \label{eik}
\end{equation}
(a comma denotes partial derivative) and the field has been decomposed into an unperturbed field and a scattered
wave of order $\Delta$.

Defining the wave four-vector $k^{ \alpha} = (\omega/c, {\bf  k})$ and setting
$$
S \equiv -k_{ \alpha} (x^{\alpha} - x_0^{\alpha}) + \Delta S =
$$
\begin{equation}
 -\omega (t-t_0) + {\bf k} \cdot ({\bf r}-{\bf r_0}) +
\Delta S, \label{Stk}
\end{equation}
it follows that the eikonal equation (\ref{eik}) implies, in a first order approximation,
\begin{equation}
\frac{\partial}{\partial t} \Delta S + c{\bf n} \cdot \nabla \Delta S + \omega (\Delta g + {\bf n} \cdot {\bf
\Delta g}) =0, \label{DS}
\end{equation}
where ${\bf k} \equiv (\omega /c) {\bf n}$ and it is understood that $k_{ \alpha} = (\omega, {\bf - k})$ .

Assuming for simplicity that the mass and velocity distributions are stationary, the Maxwell equations take the
form
\begin{equation}
{\bf \Delta \dot{E}} - i \omega ({\bf \Delta E} + {\bf n} \times {\bf \Delta H}) - c \nabla \times {\bf  \Delta H}
- i [(\omega {\bf \Delta g} + c\nabla \Delta S) \cdot {\bf E}^{(0)} ] {\bf n} = {\bf 0}
\end{equation}
\begin{equation}
{\bf \Delta \dot{H}} - i \omega ({\bf \Delta H} - {\bf n} \times {\bf \Delta E}) + c \nabla \times {\bf  \Delta E}
- i [(\omega {\bf \Delta g} + c\nabla \Delta S) \cdot {\bf H}^{(0)} ] {\bf n} = {\bf 0},
\end{equation}
where dots denote derivation with respect to time.

Setting ${\bf \Delta E}_{\bot} = (\stackrel{\leftrightarrow}{\mathbf{1}} - \stackrel{\leftrightarrow}{\mathbf{{\bf
n} {\bf n}}}) \cdot {\bf \Delta E}$ and ${\bf \Delta H}_{\bot} = (\stackrel{\leftrightarrow}{\mathbf{1}} -
\stackrel{\leftrightarrow}{\mathbf{{\bf n} {\bf n}}}) \cdot {\bf \Delta H}$, where
$(\stackrel{\leftrightarrow}{\mathbf{1}} - \stackrel{\leftrightarrow}{\mathbf{{\bf n} {\bf n}}})$ is to be
interpreted as a dyad, we obtain
\begin{equation}
\frac{\partial}{c\partial t} ({\bf \Delta E}_{\bot}  - {\bf n} \times {\bf \Delta H}_{\bot}) -
(\stackrel{\leftrightarrow}{\mathbf{1}} - \stackrel{\leftrightarrow}{\mathbf{{\bf n} {\bf n}}}) \cdot (\nabla
\times {\bf \Delta H}) - {\bf n} \times (\nabla \times {\bf \Delta E}) ={\bf 0},
\end{equation}
\begin{equation}
\frac{\partial}{c\partial t} ({\bf \Delta H}_{\bot}  + {\bf n} \times {\bf \Delta E}_{\bot}) +
(\stackrel{\leftrightarrow}{\mathbf{1}} - \stackrel{\leftrightarrow}{\mathbf{{\bf n} {\bf n}}}) \cdot (\nabla
\times {\bf \Delta E}) - {\bf n} \times (\nabla \times {\bf \Delta H}) ={\bf 0}.
\end{equation}

Also
$$
{\bf n} \cdot {\bf \Delta E} = - \Big({\bf \Delta g} + \frac{c}{\omega} \nabla \Delta S \Big) \cdot {\bf E}^{(0)},
$$
\begin{equation}
{\bf n} \cdot {\bf \Delta H} = - \Big({\bf \Delta g} + \frac{c}{\omega} \nabla \Delta S \Big) \cdot {\bf H}^{(0)},
\end{equation}
and ${\bf \Delta H}_{\bot} = {\bf n} \times  {\bf \Delta E}_{\bot}$  if terms of order $\omega^{-1}$ are
neglected. Combining all the above results, we finally arrive to the basic equation
\begin{equation}
\Big( \frac{ \partial}{c\partial t}   + {\bf n} \cdot \nabla \Big) {\bf \Delta E}_{\bot} + \Delta \gamma ~ {\bf
E}^{(0)} + \Delta p  ~{\bf n} \times {\bf E}^{(0)} ={\bf 0}, \label{pleb}
\end{equation}
where
\begin{equation}
\Delta \gamma = \frac{1}{2} \nabla \cdot \big[ (\stackrel{\leftrightarrow}{\mathbf{1}} -
\stackrel{\leftrightarrow}{\mathbf{{\bf n} {\bf n}}}) \cdot ({\bf \Delta g} + \frac{c}{\omega} \nabla \Delta
S)\big] , \label{21}
\end{equation}
\begin{equation}
\Delta p = -~ \frac{1}{2} {\bf n} \cdot (\nabla \times {\bf \Delta g}).\label{22}
\end{equation}
The first term corresponds to the absorption of the electromagnetic wave and the second term determines the
rotation of the polarization plane.

Eqs.  \eqref{DS} and \eqref{pleb} are the basic equations obtained  by Plebanski \cite{pleb}, with a different
notation. They are both of the form
\begin{equation}
 \frac{\partial}{c \partial t} F + {\bf n} \cdot \nabla  F + W (t, {\bf r})=0,\label{sol}
\end{equation}
which has the solution \cite{pleb}
\begin{equation}
F(t, {\bf r})= -  \int^t_{t_0} W (t', {\bf r} - c(t-t') {\bf n} ) dt',
\end{equation}
with the condition $F(t_0,{\bf r})=0$. This formula will be useful in what follows.

\section{Geometric optics}

It follows from Eqs. \eqref{gab} and \eqref{DS} that
\begin{equation}
 \Delta S ({\bf r},t) = \frac{2G}{c^2} \omega \int_{t_0}^t dt' \int dV' \rho({\bf r'}) \frac{\big[1 -2 {\bf
n} \cdot {\bf v} ({\bf r'})/c \big] }{|{\bf r} - c(t-t') {\bf n} - {\bf r'}|}. \label{Sint}
\end{equation}

The ray equation is given by the condition
$$
\frac{\partial S}{\partial {\bf k}}=0.
$$
Using the relations
$$
\frac{\partial \omega}{\partial {\bf k}} = c{\bf n},
$$
$$
\frac{\partial}{\partial {\bf k}}= \frac{c}{\omega} (\stackrel{\leftrightarrow}{\mathbf{1}} -
\stackrel{\leftrightarrow}{\mathbf{{\bf n} {\bf n}}}) \cdot \frac{\partial}{\partial {\bf n}},
$$
together with \eqref{Stk}, it follows that the trajectory is given by
\begin{equation}
{\bf r} (t) = {\bf r}_0 + c(t-t_0) {\bf n} - \frac{c}{\omega} {\bf n} \Delta S - \frac{c}{\omega}
(\stackrel{\leftrightarrow}{\mathbf{1}} - \stackrel{\leftrightarrow}{\mathbf{{\bf n} {\bf n}}}) \cdot
\frac{\partial}{\partial {\bf n}} \Delta S,\label{rt}
\end{equation}
with  $\Delta S$  given by \eqref{Sint}.

The unperturbed trajectory is simply
$$
{\bf r} (t) = {\bf r_0} + c(t-t_0) {\bf n},
$$
where of course ${\bf r}_0= {\bf r}(t_0)$, and this, in a first order approximation,  can be substituted in Eq.
\eqref{rt} after using the explicit form of $\Delta S$ given by Eq. \eqref{Sint} and taking into account that
$$
{\bf r} - c(t-t') {\bf n} - {\bf r'} = {\bf r_0} + c(t'-t_0) {\bf n} - {\bf r'}.
$$
It then follows that the \emph{unit} vector in the direction of propagation
$${\bf u}(t) \propto \frac{d {\bf r}}{dt}(t)$$
is
$$
{\bf u}(t) = {\bf n} -\frac{2G}{c^2}(\stackrel{\leftrightarrow}{\mathbf{1}} - {\bf n} {\bf n}) \cdot \int dV'
\rho({\bf r'}) \Big\{ -   \frac{2{\bf v} ({\bf r'})}{c|{\bf r_0} - {\bf r'} +c(t-t_0) {\bf n}| } +
$$
\begin{equation}
\big[1-  2 {\bf n} \cdot {\bf v} ({\bf r'})/c  \big] c\int_{t_0}^t dt'
  \frac{{\bf r_0} -{\bf r'} }{|{\bf r_0} - {\bf r'} + c(t'-t_0) {\bf n}
|^3} \Big\} .\label{ut}
\end{equation}

Furthermore,  ${\bf u}$ (or ${\bf r}_0$) admits a direct decomposition into a direction parallel to ${\bf n}$ and
one perpendicular to ${\bf n}$, which we can choose as ${\bf b} \equiv (\stackrel{\leftrightarrow}{\mathbf{1}} -
\stackrel{\leftrightarrow}{\mathbf{{\bf n} {\bf n}}})\cdot {\bf r}_0$, where ${\bf b}$ can be interpreted as the
impact parameter of the light-ray. A similar decomposition for the initial position ${\bf r}_0$ can be performed:
since ${\bf r}_0= {\bf r}(t_0)$, it follows that in the limit $t_0 \rightarrow -\infty$ we have ${\bf r}_0 - {\bf
b} \rightarrow t_0 {\bf n}$. We can then set $t_0 \rightarrow -\infty$ and take
$$
{\bf r_0}  + c(t'-t_0) {\bf n} = {\bf b} +ct' ~{\bf n}
$$
in the above Eq. (\ref{ut}). It thus follows that the vector ${\bf u}$ changes from ${\bf u}_{in}={\bf n}$ at $t
\rightarrow -\infty$ to ${\bf u}_{out}$ at $t \rightarrow \infty$ as
\begin{eqnarray}
{\bf u}_{out}& =& {\bf u}_{in} - \\ \nonumber
 &&\frac{4G}{c^2}(\stackrel{\leftrightarrow}{\mathbf{1}} -
\stackrel{\leftrightarrow}{\mathbf{{\bf n} {\bf n}}}) \cdot \int dV \rho({\bf r}) \big[1- 2 {\bf n} \cdot {\bf v}
({\bf r})/c  \big]
  \frac{{\bf b} -{\bf r} }{({\bf b} - {\bf r})^2 - ({\bf n}\cdot {\bf r})^2 }. \label{out2}
\end{eqnarray}

For a rotating mass configuration, the velocity is given in terms of the angular velocity vector ${\bf \Omega}$ as
${\bf v} = {\bf \Omega} \times {\bf r}$. We choose a Cartesian coordinate system such that  ${\bf \Omega}$ is in
the $z$ direction and ${\bf n}$ in the $(y,z)$ plane forming an angle $\alpha$ with ${\bf \Omega}$. Thus, without
loss of generality,
$$
{\bf \Omega} = \Omega ({\bf r}) (0,0,1),
$$
$$
{\bf n} = (0, \sin \alpha, \cos \alpha),
$$
\begin{equation}
{\bf b}  \equiv  (b_x, -\beta \cos \alpha ,\beta \sin \alpha) = b_x {\bf e_x} -\beta {\bf n} \times {\bf e}_x
.\label{bbb}
\end{equation}
Defining a complex basis vector
\begin{equation}
 \boldsymbol\epsilon =  {\bf e}_x + i {\bf n} \times {\bf e_x}~,\label{ep}
\end{equation}
together with its complex conjugate $\boldsymbol\epsilon^*$, it follows that the  equation for the deflection of
light can be written in the compact form
\begin{equation}
{\bf u}_{out} = {\bf u}_{in} - \Delta u ~\boldsymbol\epsilon - (\Delta u ~\boldsymbol\epsilon)^*,\label{uc}
\end{equation} where
\begin{equation}
 \Delta u = \frac{2G}{c^2} \int dV \rho({\bf r})  \frac{\big[1- 2 c^{-1} \sin \alpha ~{\Omega}({\bf r})~ x  \big]
 }{b_x -x-i (\beta + \cos \alpha ~y -\sin \alpha ~z)} . \label{utout2xy}
\end{equation}

\section{Vectorial optics}

Let us return to the basic equation \eqref{pleb}. Since it is of the form \eqref{sol}, its solution is
\begin{equation}
\Delta {\bf E}_{\bot} = \Delta \Gamma ~{\bf E}^{(0)} + \Delta P ~{\bf n} \times {\bf E}^{(0)},
\end{equation}
where
\begin{equation}
\Delta  \Gamma  = -c\int^t_{-\infty} dt' ~\Delta \gamma (t', {\bf r}- c(t-t') {\bf n} ),
\end{equation}
\begin{equation}
\Delta  P  = -c \int^t_{-\infty} dt' ~\Delta p(t', {\bf r}- c(t-t') {\bf n} ),
\end{equation}
and $\Delta  \gamma$ and $\Delta  p$ are to be calculated from Eqs. \eqref{21} and \eqref{22}. For this purpose,
we first evaluate the term appearing in Eq. \eqref{21}:
$$
\nabla \cdot (\stackrel{\leftrightarrow}{\mathbf{1}} - \stackrel{\leftrightarrow}{\mathbf{{\bf n} {\bf n}}}) \cdot
\nabla \Delta S \equiv   \nabla^2_{\bot}\Delta S .
$$
Setting ${\bf R} ={\bf r} - {\bf r}' -c(t-t') {\bf n}$ and using the relation
$$
\nabla^2_{\bot} R^{-1} = \frac{\partial}{c\partial t'} \Big( \frac{c(t'-t) + {\bf n} \cdot ({\bf r} - {\bf r}')
}{R^3}\Big),
$$
it follows directly from Eq. \eqref{Sint} that
\begin{equation} \omega^{-1} \nabla^2_{\bot} \Delta S =
 \frac{2G}{c^3} \int dV' \rho ({\bf r'}) [1-2 {\bf n} \cdot {\bf v} ({\bf r'})/c ] \frac{{\bf n} \cdot ({\bf
r}-{\bf r'})}{|{\bf r}-{\bf r'}|^3}.
\end{equation}

Using also the formula \eqref{gab} for the vector ${\bf \Delta g}$, we obtain
\begin{equation}
\Delta \gamma = \frac{G}{c^2} \int dV' \rho ({\bf r'}) [{\bf n} -2 {\bf v} ({\bf r'})/c ] \cdot \frac{({\bf
r}-{\bf r'})}{|{\bf r}-{\bf r'}|^3}
\end{equation}
and
\begin{equation}
\Delta p = \frac{2G}{c^3} \int dV' \rho ({\bf r'}) [ {\bf v} ({\bf r'}) \times {\bf n} ] \cdot \frac{({\bf r}-{\bf
r'})}{|{\bf r}-{\bf r'}|^3}.
\end{equation}

Now, in order to calculate $\Delta \Gamma$ and $ \Delta P$, we use the fact that an integral of the form
\begin{equation}
I \equiv \int_{-\infty}^t dt' \frac{{\bf r}-{\bf r'} - c(t-t'){\bf n}}{|{\bf r}-{\bf r'} - c(t-t'){\bf n}|^3}
\end{equation}
can be approximated, following the same procedure as in the previous section, by
\begin{equation} I \approx
\int_{-\infty}^{\infty} dt' \frac{{\bf b}-{\bf r'}+ct'~{\bf n}}{\big[c^2 t^{'2} - 2 {\bf n} \cdot {\bf r'}~ ct' +
({\bf b}-{\bf r'})^2\big]^{3/2}} = \frac{2}{c} ~\frac{{\bf b}-{\bf r'}+({\bf n} \cdot {\bf r'})~ {\bf n}}{ ({\bf
b}-{\bf r'})^2 -({\bf n} \cdot {\bf r'})^2}
\end{equation}

It thus follows that
\begin{equation} \Delta \Gamma = \frac{4G}{c^3} \int dV ~ \rho ({\bf r})  {\bf v} ({\bf r})
\cdot \frac{{\bf b}-{\bf r}+({\bf n} \cdot {\bf r})~ {\bf n}}{ ({\bf b}-{\bf r})^2 -({\bf n} \cdot {\bf r})^2}
\end{equation}
and
\begin{equation}
\Delta P = \frac{4G}{c^3} \int dV ~\rho ({\bf r}) [{\bf n}  \times {\bf v} ({\bf r}) ] \cdot \frac{{\bf b}-{\bf
r}}{({\bf b}-{\bf r})^2 -({\bf n} \cdot {\bf r})^2}.
\end{equation}

As in the previous section, we can rewrite these formulas in the more compact form as
\begin{equation}
\Delta P  +i \Delta \Gamma = - \frac{2G}{c^3} \int dV~ \rho ({\bf r}) ~\Omega ({\bf r}) ~\frac{\cos \alpha ~ x
+iy}{b_x -x -i (\beta + \cos \alpha ~y - \sin \alpha ~z)}  ~,
\end{equation}
as can be checked with a little algebra.

\section{Axial symmetry}

The above formulas can be simplified for axisymmetric configurations in which the density $\rho$ and the velocity
${\bf v}$ do not depend on the azimuthal angle $\phi$ around the symmetry axis. In this case, the integration over
$\phi$ can be performed in general using the formulas given in the Appendix for the integrals defined there as
${\cal I}_n$ ($n=, 1,2,3$).

As shown in the Appendix, it is convenient to use oblate spheroidal coordinates $(u,\eta)$, or equivalently $v
\equiv \cos \eta$, as defined by Eqs. \eqref{obl}. Let us start with Eq. \eqref{utout2xy}, which has two terms:
one dependent and one independent of the angular speed $\Omega$. Accordingly, we set $ \Delta u \equiv \Delta u_1
+ \Delta u_2$, where
$$
\Delta u_1 = \frac{2G}{c^2} a^3 \int_0^{\infty} du \int_{-1}^1 dv ~(u^2 + v^2) \rho(u,v)~{\cal I}_1(u,v)
$$
\begin{equation}
=i ~\frac{4\pi G}{c^2 b_x} a^3 \int_0^{\infty} du  \Big\{  \int_{-1}^{-\cos \alpha}dv -\int_{\cos \alpha}^{1}dv~
\Big\} (u + iv) \rho(u,v)
\end{equation}
and
$$
\Delta u_2 = -\frac{4G}{c^3} a^3 \sin \alpha \int_0^{\infty} du \int_{-1}^1 dv~ (u^2 + v^2) \rho(u,v)   ~\Omega
(u,v) ~{\cal I}_3 (u,v)
$$
$$
 = \frac{8\pi G}{c^3 \sin \alpha} a^3 \int_0^{\infty} du
 \Big\{ -i (1+v) \int_{-1}^{-\cos \alpha}dv ~(1+iu)(u+iv)
$$
$$
 + (1-\cos \alpha) \int_{-\cos \alpha}^{\cos \alpha}dv ~(u^2 +v^2)
$$
\begin{equation}
+i(1-v) \int_{\cos \alpha}^{1} dv ~(1-iu)(u+iv) \Big\}
  \rho(u,v)  ~\Omega
(u,v) .
\end{equation}

Similarly we have
$$
\Delta P - i \Delta \Gamma = - \frac{2 G}{c^3} a^3 \int_0^{\infty} du \int_{-1}^1 dv ~(u^2 + v^2) \rho(u,v) \Omega
(u,v) ~{\cal I}_2
$$
and thus
\begin{equation}
\Delta P = \frac{4 \pi G}{c^3} a^3 \int_0^{\infty} du \int_{-\cos \alpha}^{\cos \alpha} dv ~(u^2 + v^2) \rho(u,v)
\Omega (u,v),
\end{equation}
\begin{equation}
\Delta \Gamma =0.
\end{equation}
There is no absorption.

\subsection{Application}

As an example of application, consider a massive configuration described by a spherically symmetric Plummer
potential \cite{plum}
\begin{equation}
\psi = - \frac{GM}{(r^2 + \overline{\epsilon})^{1/2}},
\end{equation}
where $M$ is the total mass of the configuration and $\overline{\epsilon}$ is a certain parameter. The mass
density, given through the Poisson equation, is
\begin{equation}
\rho =  \frac{3M}{4\pi} ~ \frac{\overline{\epsilon}^2}{(r^2 + \overline{\epsilon})^{5/2}}.
\end{equation}
In cylindrical-ellipsoidal coordinates, as defined in the Appendix, we have
\begin{equation}
r^2 = R^2 + z^2 = a^2 \big( 1 + u^2 -v^2 + 2 \lambda uv +\lambda^2 +\epsilon^2 \big),
\end{equation}
where $\lambda \equiv \beta /b_x$  and $ \epsilon \equiv   \overline{\epsilon}/a = (\overline{\epsilon}/b_x) \sin
\alpha $.

It follows with some lengthy but straightforward algebra that
$$
\Delta u \approx \Delta u_1 = \frac{2GM}{c^2 b_x (1-i \lambda)} \Big[ \frac{(\lambda ^2 + 1)\sin^2
\alpha}{(\lambda ^2 + 1)\sin^2 \alpha ~+ \epsilon^2}~ -
$$
\begin{equation}
i \frac{\epsilon^2}{1+\lambda^2 +\epsilon^2} \Big( \frac{1}{\sqrt{\lambda^2 + \epsilon^2}} -\frac{\cos
\alpha}{\sqrt{\lambda^2 + \epsilon^2 + \sin^2 \alpha}} \Big)  \Big],
\end{equation}
neglecting the term $\Delta u_2$ which is of order in $(v/c) \Delta u_1$. More explicitly,
$$
\Delta u_1 = \frac{2GM}{c^2 (b_x -i \beta)} \Big[ \frac{b^2}{b^2 + \overline{\epsilon}^2}~ -
$$
\begin{equation}
i \frac{\overline{\epsilon}^2 b_x \sin^2 \alpha}{b^2 +\overline{\epsilon}^2 \sin^2 \alpha} \Big(
\frac{1}{\sqrt{\beta^2 + \overline{\epsilon}^2 \sin^2 \alpha}} -\frac{\cos \alpha}{\sqrt{\beta^2 + (
\overline{\epsilon}^2 + b_x^2 )\sin^2 \alpha}} \Big) \Big],
\end{equation}
where $b = \sqrt{b_x^2 + \beta^2}  $. The deviation of a light ray follows from Eq. \eqref{uc}.

 For the particular case of a rigid rotation, with constant $\Omega$, it also follows that
\begin{equation}
\Delta P = \frac{2GM}{c^3 } ~ \frac{\overline{\epsilon}^2 }{b^2 + \overline{\epsilon}^2}~\Omega~\cos \alpha .
\end{equation}

In the limit $\overline{\epsilon} \rightarrow 0$, we recover using Eq. \eqref{uc} the usual formula for the
deflection of light by a point mass $M$:
\begin{equation}
{\bf u}_{out} = {\bf n} - \frac{4 G M}{c^2 b^2} {\bf b},
\end{equation}
recalling that  ${\bf b}$ is the impact parameter vector. The polarization is not altered in this particular case
in which light propagates in vacuum: $\Delta P =0$.

\section{Concluding remarks}

We have obtained formulas in closed form for the deflection, absorption and rotation of the polarization plane of
an electromagnetic wave, as produced by a massive rotating cloud interacting gravitationally with light. We
illustrated the use of these formulas with a particular model that yields analytic solutions. Even for this simple
model, the solutions are somewhat cumbersome, but it must be relatively easy to perform numerical calculations
with the given formulas for physically realistic distributions of mass and angular velocity. The distribution
function of a self-gravitating cloud cannot be deduced on purely theoretical grounds, since the collisionless
Boltzman equation yields only certain restrictions on it \cite{bintrem}, but there are several observational
indications of how this distribution should be (see, e.g., \cite{tn,ea}). The formalism of the present article can
be applied to any model of mass and angular momentum distribution.


\section*{Appendix A}

\renewcommand{\theequation}{\Alph{section}.\arabic{equation}}

\setcounter{section}{1} \setcounter{equation}{0}

Consider the integral
\begin{equation}
{\cal I}_1 = \int^{2\pi}_0 d \phi ~\frac{1}{b_x -x -i(\beta + \cos \alpha ~y - \sin \alpha ~z) },
\end{equation}
where $x=R \cos \phi$ and $y= R \sin \phi$ in cylindrical coordinates $(R, \phi, z)$. It can be expressed as a
contour integral in the complex $Z=e^{i\phi} $ plane:
$$
{\cal I}_1 =  \oint dZ~ F_1(Z),
$$
where the integration is along the unit circle $|Z|^2 =1$ and
\begin{equation}
F_1(Z) =   \frac{2i }{ (1 + \cos \alpha)R Z^2 - 2  (b_x - i \beta + i \sin \alpha ~z) Z + (1 - \cos
\alpha)R}.\label{I}
\end{equation}

It can be seen that $F_1(Z)$ has two poles located at
$$
Z = Z_{\pm} \equiv \tan \Big(\frac{ \alpha}{2} \Big) \exp\{\pm i \Phi\},
$$
where $\Phi$ is a complex function defined as
$$
\frac{b_x - i \beta}{\sin \alpha} + iz = R \cos \Phi.
$$
The residues at each poles are then
\begin{equation}
{\rm lim}_{Z \rightarrow Z_{\pm}} [(Z-Z_{\pm})F_1(Z)] = \pm (R \sin \alpha \sin \Phi )^{-1}. \label{resi}
\end{equation}

At this point it is convenient to define oblate ellipsoidal coordinates $(u, \eta)$ such that
\begin{equation}
R = a \sqrt{u^2 +1} ~\sin \eta, \quad z= a u \cos \eta + \frac{\beta}{\sin \alpha} ,\label{obl} \end{equation}
where $a \equiv b_x /\sin \alpha$, and $u \geq 0$ and $0 \leq \eta < \pi$.

Accordingly
$$
 R \sin \Phi = a (u - i \cos \eta)
 $$
 $$
  R \cos \Phi = a (1 + i u   \cos \eta)
 $$
 or equivalently
$$
 \exp\{- i \Phi\} = \sqrt{\frac{1-iu}{1+iu}}  ~\tan \Big(\frac{\eta}{2}\Big).
$$
The residues at the poles, according to \eqref{resi}, are
$$
 \pm \frac{1}{b_x (u - i \cos \eta)}
$$

We must now distinguish three cases that determine which poles are inside the unit circle. For this purpose, note
that
$$
|Z_+| = \tan \Big(\frac{\alpha}{2} \Big) \Big/ \tan (\frac{\eta}{2} \Big)
$$
$$
|Z_-| = \tan \Big(\frac{\alpha}{2} \Big)  \tan (\frac{\eta}{2} \Big),
$$
and therefore we have

i)~ $\eta < \alpha$: one inner pole at $Z=Z_-$.

ii)~$\eta > \pi - \alpha$: one inner  pole at $Z=Z_+$.

iii)~ $\alpha < \eta < \pi -  \alpha$: two inner poles at  $Z=Z_+$ and $Z=Z_-$.

Given the residues at the poles it then follows that
\begin{equation}
{\cal I}_1 =  \frac{2 \pi i}{b_x (u - i \cos \eta)}
\begin{cases}
  -1 & \text{if $\eta <\alpha$}, \\
    ~ 0 & \text{if $\alpha <\eta < \pi - \alpha$}, \\
  ~ 1 & \text{if $\pi - \alpha <\eta $}.
  \end{cases}
\end{equation}

Another integral appearing in the text is
\begin{equation}
{\cal I}_2 = \int^{2\pi}_0 d \phi ~\frac{\cos \alpha~ x  ~+iy}{b_x -x - i(\beta + \cos \alpha ~y - \sin \alpha
~z)}.
\end{equation}
Since the integrand is an exact derivative with respect of $\phi$, it  follows that ${\cal I}_2 = 2\pi N$, where
$N$ is a winding number that depends on the parameters in the integrand. In order to determine this number,we also
express the above integral as a contour integral in the complex $Z$ plane:
$$
{\cal I}_2 =  \oint dZ~ F_2(Z),
$$
where the integration is along the unit circle $|Z|^2 =1$ and
\begin{equation}
F_2(Z) =   \frac{i}{Z }~ \frac{(1 + \cos \alpha) Z^2 - 1 + \cos \alpha }{ (1 + \cos \alpha) Z^2 - 2 R^{-1} (b_x -i
\beta + i \sin \alpha ~z) Z + 1 - \cos \alpha}.\label{H}
\end{equation}

There are three poles as in the previous cases, and the residues are
$$
{\rm lim}_{Z \rightarrow 0} [Z~F_2(Z)] =-i,
$$
$$
{\rm lim}_{Z \rightarrow Z_{\pm}} [(Z-Z_{\pm})~F_2(Z)] =i.
$$
Accordingly,
\begin{equation}
{\cal I}_2 =
\begin{cases}
  ~~0 & \text{if $\eta <\alpha$}, \\
   -2 \pi & \text{if $\alpha <\eta < \pi - \alpha$}, \\
  ~~ 0 & \text{if $\pi - \alpha <\eta $}.
  \end{cases}
\end{equation}

Yet another integral that appears in the text is
\begin{equation}
{\cal I}_3 = \int^{2\pi}_0 d \phi ~\frac{x}{b_x -x -i(\beta + \cos \alpha ~y - \sin \alpha ~z) }~.
\end{equation}
Since
$$
\sin^2 \alpha ~ {\cal I}_3 = (b_x -i \beta  +i \sin \alpha ~z)~ {\cal I}_1 - \cos \alpha~ {\cal I}_2 - 2 \pi~,
$$
it follows that
\begin{equation}
  {\cal I}_3 =-~\frac{2 \pi}{\sin^2 \alpha}
  \begin{cases}
   i(1-iu)(1-\cos \eta)\big/(u-i \cos \eta) & \text{if $\eta <\alpha$}, \\
    1-\cos \alpha & \text{if $\alpha <\eta < \pi - \alpha$}, \\
   -i(1+iu)(1+\cos \eta)\big/(u-i \cos \eta) & \text{if $\pi - \alpha <\eta $}.
  \end{cases}
\end{equation}

\end{document}